\documentstyle[epsf]{article}
\begin{document}
\renewcommand{\thefootnote}{\fnsymbol{footnote}}
\begin{center}
{\bf \large
A Measurement of the $D^{*\pm}$ Cross Section in Two-Photon Processes
\footnote{
published in Phys. Rev. {\bf D 50}, 1879 (1994).
}
}
\vskip 0.5cm
TOPAZ Collaboration\\
\vskip 0.5cm
R.Enomoto$^{(1)}$\footnote{internet address: enomoto@kekvax.kek.jp},
M.Iwasaki$^{(2)}$,
K.Muramatsu$^{(2)}$,
H.Hayashii$^{(2)}$,
A.Miyamoto$^{(1)}$,
R.Itoh$^{(1)}$,
K.Abe$^{(3)}$,
T.Abe$^{(4)}$,
I.Adachi$^{(1)}$,
M.Aoki$^{(4)}$,
S.Awa$^{(2)}$
R.Belusevic$^{(1)}$,
K.Emi$^{(3)}$,
H.Fujii$^{(1)}$,
K.Fujii$^{(1)}$,
T.Fujii$^{(5)}$,
J.Fujimoto$^{(1)}$,
K.Fujita$^{(6)}$,
N.Fujiwara$^{(2)}$,
B.Howell$^{(7)}$,
N.Iida$^{(1)}$,
H.Ikeda$^{(1)}$,
H.Iwasaki$^{(1)}$,
R.Kajikawa$^{(4)}$,
S.Kato$^{(8)}$,
S.Kawabata$^{(1)}$,
H.Kichimi$^{(1)}$,
M.Kobayashi$^{(1)}$,
D.Koltick$^{(7)}$,
I.Levine$^{(7)}$,
K.Miyabayashi$^{(4)}$,
K.Nagai$^{(9)}$,
T.Nagira$^{(2)}$,
E.Nakano$^{(4)}$,
K.Nakabayashi$^{(4)}$,
O.Nitoh$^{(3)}$,
S.Noguchi$^{(2)}$,
F.Ochiai$^{(10)}$,
Y.Ohnishi$^{(4)}$,
H.Okuno$^{(8)}$,
T.Okusawa$^{(6)}$,
K.Shimozawa$^{(4)}$,
T.Shinohara$^{(3)}$,
A.Sugiyama$^{(4)}$,
N.Sugiyama$^{(11)}$,
S.Suzuki$^{(4)}$,
K.Takahashi$^{(3)}$,
T.Takahashi$^{(6)}$,
M.Takemoto$^{(2)}$,
T.Tanimori$^{(11)}$,
T.Tauchi$^{(1)}$,
F.Teramae$^{(4)}$,
Y.Teramoto$^{(6)}$,
N.Toomi$^{(2)}$,
T.Toyama$^{(4)}$,
T.Tsukamoto$^{(1)}$,
S.Uno$^{(1)}$,
Y.Watanabe$^{(11)}$,
A.Yamaguchi$^{(2)}$,
A.Yamamoto$^{(1)}$, and
M.Yamauchi$^{(1)}$\\
\vskip 0.5cm
{\it
$^{(1)}$National Laboratory for High Energy Physics, KEK,
  Ibaraki-ken 305, Japan
\\
$^{(2)}$Department of Physics, Nara Women's University,
 Nara 630, Japan
\\
$^{(3)}$Dept. of Applied Physics,
Tokyo Univ. of Agriculture and Technology,
 Tokyo 184, Japan
\\
$^{(4)}$Department of Physics, Nagoya University,
 Nagoya 464, Japan
\\
$^{(5)}$Department of Physics, University of Tokyo,
   Tokyo 113, Japan
\\
$^{(6)}$Department of Physics, Osaka City University,
 Osaka 558, Japan
\\
$^{(7)}$Department of Physics, Purdue University,
 West Lafayette, IN 47907, USA
\\
$^{(8)}$Institute for Nuclear Study, University of Tokyo,
   Tokyo 188, Japan
\\
$^{(9)}$The Graduate School of Science and Technology,
Kobe University,
Kobe 657, Japan
\\
$^{(10)}$Faculty of Liberal Arts, Tezukayama Gakuin University,
 Nara 631, Japan
\\
$^{(11)}$Department of Physics, Tokyo Institute of Technology,
     Tokyo 152, Japan
}
\end{center}
\begin{abstract}
We have measured the inclusive $D^{*\pm}$ production cross section
in a two-photon collision at the TRISTAN $e^+e^-$ collider. The
mean $\sqrt{s}$ of the collider
was 57.16 GeV and the integrated luminosity was
150 $pb^{-1}$.
The differential cross section ($d\sigma(D^{*\pm})/dP_T$)
was obtained in the $P_T$ range between 1.6 and 6.6 GeV and
compared with theoretical predictions, such as
those involving direct and resolved photon processes.
\end{abstract}

Hadron production in two-photon collisions is
described by the vector-meson dominance model,
the quark-parton model (direct process) \cite{r1}, and the hard scattering
of the hadronic constituents of almost-real
photons (resolved photon process)
\cite{r2,r3,r4,r5}, which has been observed by the
previous experiments
\cite{r6}.
However, more detailed studies are necessary in order to
understand these processes quantitatively.
Heavy quark pair production processes are good probes, since
the theoretical calculations are less ambiguous
than for light quarks\cite{r23}.

The previous measurements had been carried out at around
$\sqrt{s}\sim 30~GeV$ \cite{r7,r8,r9}, and
are consistent with a recent theoretical prediction\cite{r23}.
At the TRISTAN energy ($\sqrt{s}\sim$ 60 GeV),
the $c\bar{c}$ production cross section becomes
sizable; we have obtained the largest statistics for
this type of process.
We carried out a measurement
of the $D^{*\pm}$ production cross section at a $P_T$ greater
than 1.6 GeV in two-photon collision events.

The detail concerning the TOPAZ detector can be found elsewhere \cite{r10}.
The integrated luminosity of the event sample used in the analysis
was 150 $pb^{-1}$. The mean $< \sqrt{s}>$ of the collider
was 57.16 GeV.
The trigger conditions are as follows:
more than two tracks with $P_T>0.3 \sim 0.7$ GeV
and opening angle $>$ 45$\sim$70 degrees (depending on the beam
condition); a neutral energy deposit in the barrel calorimeter
be greater than
4 GeV; or that in the endcap calorimeters be greater than 10 GeV.

The selection criteria for two-photon events are as follows:
the number of charged tracks be $\geq$ 4;
the total visible
energy in the central part of the detector be between
4 and 25 GeV; the vector sum of the transverse momenta of
the particles with respect to the beam axis
($|\Sigma \vec{P}_T|$) be less than 7.5 GeV;
the sum of the charges be $\leq$ 3; the event vertex be consistent
with the
beam crossing point;
and no large energy clusters ($E>0.25E_{beam}$) in the
barrel calorimeter.
In addition, we divided each event into two jets with respect to the
plane perpendicular to the thrust axis,
and the cosine of the angle between
the two jets was required to be greater than
-0.9. These restrictions were made in order
to reduce beam-gas, and single-photon-exchange
events. A total of 10788 events was selected.

The charged track selections were as follows: the closest approach
to the event vertex be consistent within the measurement error;
the number of degrees of freedom (DOF)
in the track-fitting be $\geq$ 3; and $P_T$
be $\geq$ 0.15 GeV.

The $\gamma$ selections were: the cluster be detected by a
barrel-type
lead-glass calorimeter; the energy be $\geq$ 0.2 GeV; and the
cluster position be separated from any charged-track extrapolations.
In addition, the $e^+e^-$ pairs which were consistent with
$\gamma$ conversion at the inner vessel of the Time Projection
Chamber (TPC) were reconstructed (1C-fit) and used as $\gamma$
candidates.

In order to reconstruct charm quarks, we used the decay modes
$D^{*\pm}\rightarrow \pi^{\pm}D^0(\bar{D^0})$, followed by
$(D^0(\bar{D^0})\rightarrow K^{\mp}\pi^{\pm}X(\bar{X}))$.
{}From now on, any mention of decays includes the charge conjugation
modes, for simplicity. For $D^0$ decays, the decay
modes $D^0\rightarrow K^-\pi^+$,
$K^-\pi^+\rho^0$,
$K^-\pi^+\pi^0$ ($K^{*-}\pi^+$, $K^-\rho^0$,
$\bar{K}^{*0}\pi^0$), and
$K^-\pi^+\pi^0\pi^0$ were reconstructed
by using kinematical constraint fits (1-3C).
The cuts on $\chi^2_{fit}$ were required to be greater than the 5\%
confidence level (CL).
These decay modes were selected because they have
relatively high acceptances,
considering the branching fractions and detector acceptances.
For more than two-body decay modes,
we applied a dE/dx cut in selecting $K^-$;
for a two-body decay, a sufficient S/N was obtained without
this cut.
The cuts on the vector meson masses were carried out according
to the detector resolution and their intrinsic decay widths.
In the case of vector-plus-pseudo-scalar decay, we applied
cut on the angle of the decay product of the vector
meson $(\theta_{VP})$ in its
center-of-mass frame with respect to the vector
meson line of flight, i.e., $|cos\theta_{VP}|>0.5$.

The $D^{*^+}$s were reconstructed with those $D^0$ candidates
mentioned above while combining $\pi^+$s (soft-pions from
hereafter) with momenta less than
0.65 GeV. The energy fraction, $z=E(D^{*+})/E_{beam}$, was required
to be between 0.1 and 0.25. We then calculated the mass differences,
i.e., $dM=M(\pi^+D^0)-M(D^0)$.
The dM distribution is plotted in Fig.1(a).
\begin{figure}
\epsfysize15cm
\hskip1in\epsfbox{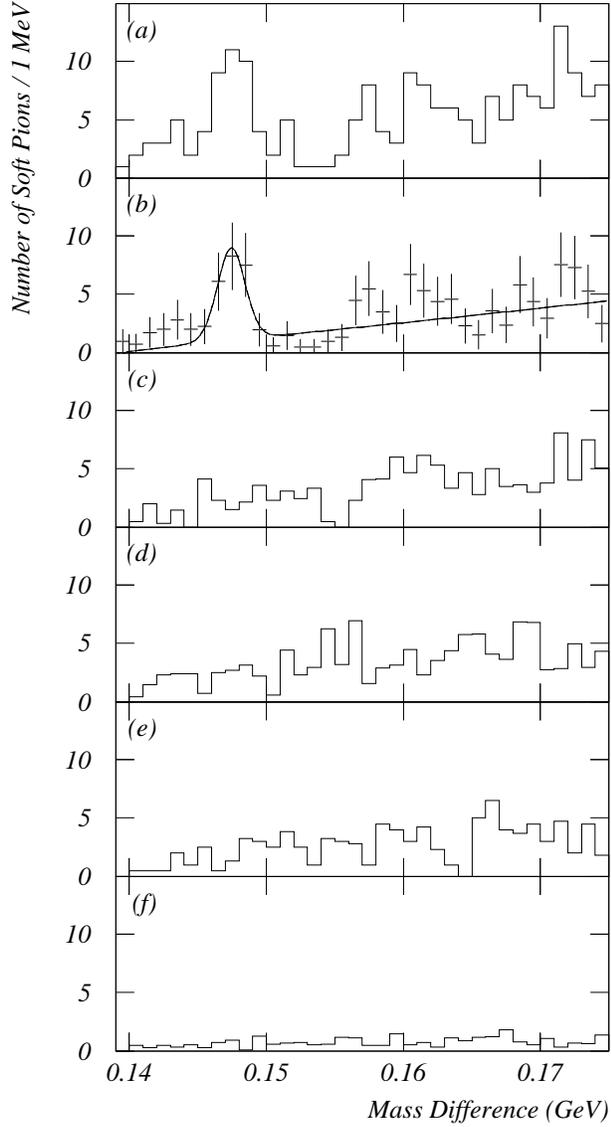}
\caption{
Mass-difference ($dM=M(D^0\pi^\pm)-M(D^0))$ distributions:
(a) dM distribution without the weighting; (b) after the
weighting. The curve is obtained from a best-fit
function described in the text;
(c) dM distribution
for wrong-sign soft-pion; (d) that for wrong-sign kaon; (e) that
for wrong-sign soft-pion and wrong-sign kaon; and (f) background
estimation from $e^+e^-\rightarrow \gamma \rightarrow q\bar{q}$
processes.
}
\label{fig1}
\end{figure}

\begin{figure}
\epsfysize7cm
\hskip1in\epsfbox{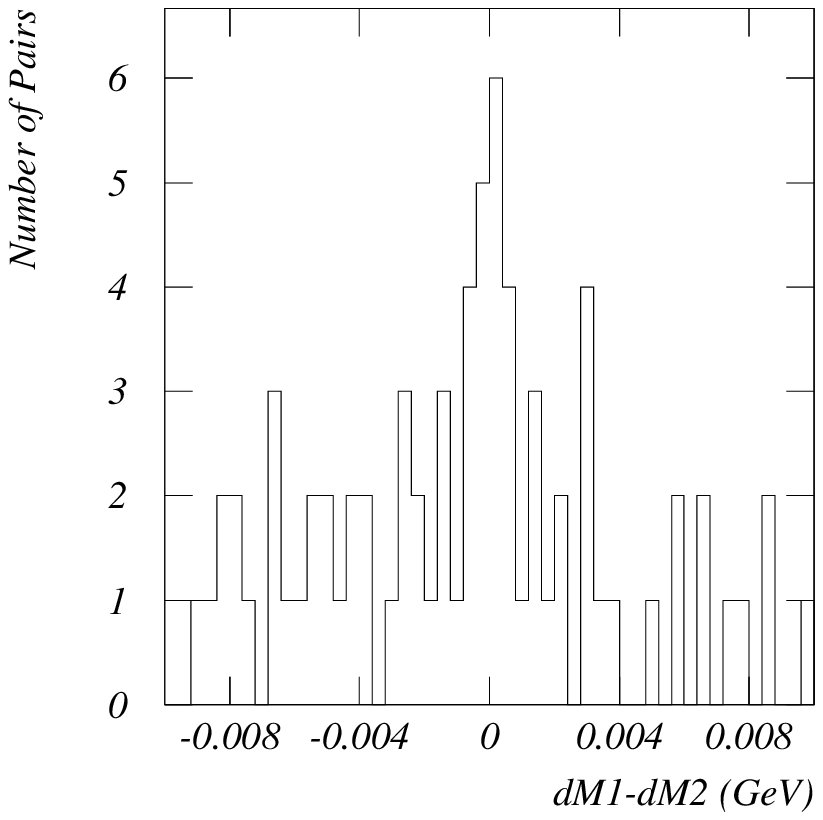}
\caption{
Differences of dM values for pairs sharing the same soft-pions
in the same events.
}
\label{fig2}
\end{figure}

\begin{figure}
\epsfysize15cm
\hskip1in\epsfbox{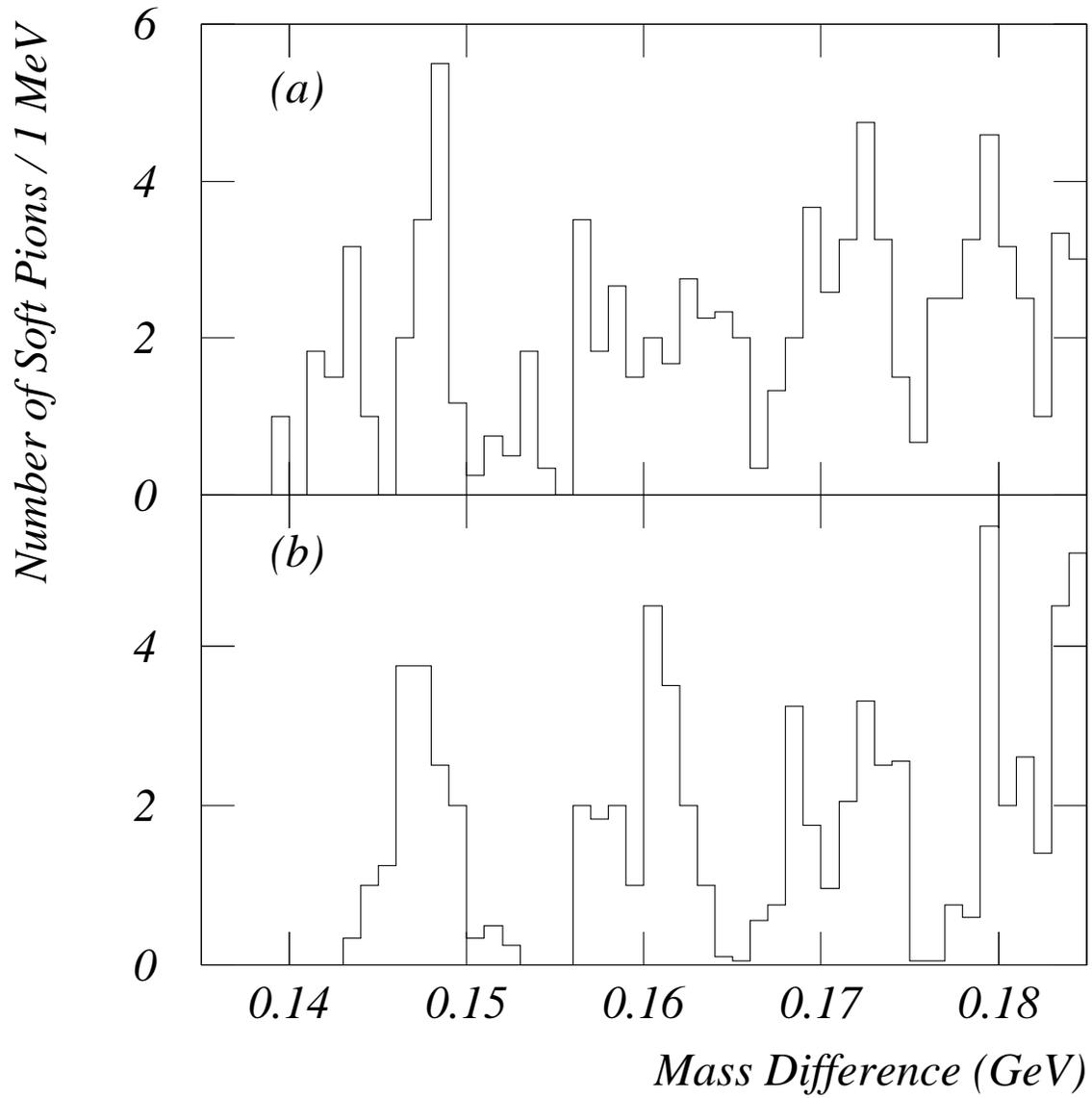}
\caption{Mass-differences of two cases: (a) when the soft-pion
passed near the TPC support structures, and (b) when it passed
far away from them.}
\label{fig3}
\end{figure}

There were multi-$D^0$-candidates for one soft-pion which gave
similar $dM$s. These occurred when one of the lowest
momentum particles of the $D^0$ decay
was misidentified\cite{r11}.
In these cases, each entry gave similar dM value. The differences
of dM values for pairs sharing the same soft-pions
in the same events were plotted in Fig.\ \ref{fig2}.
The peak around zero caused an overestimate of the statistical
significance of the signal.
In order to cure them, we carried out a
weighting method, i.e., when there was more than one $D^0$-candidate
for a given soft-pion, each entry in the $dM$ histogram was
weighted by the reciprocal of the number of candidates.
By this procedure, we could avoid any overestimation of the statistical
significance of the peak entry. This was checked using a Monte-Carlo
simulation.
The
resulting $dM$ distribution is shown in Fig.\ \ref{fig1}-(b).
The mean visible energy at the rest frame (WVIS) for the events
 containing the $D^{*\pm}$ candidates were 5.3 GeV.
In order to check the peak around the $D^{*\pm}$ region,
we carried out the wrong-sign combination such as $D^0\pi^-$ and etc.
The results are plotted in Fig.\ \ref{fig1}-(c), (d), and (e). There were no
peak structures at all.
The excess below the $D^{*\pm}$ peak was explained by
the energy loss at
support structures of
the inner field cage
and the central membrane
of the TPC, which were
distributed inhomogeneously,
whereas the correction for them were made only in average.
Figs \ref{fig3} show the mass-differences of two cases:
(a) when the soft-pion passed near
these supports, and (b) when it passed far away from them.
This happened because the soft-pions in the two photon events
had extremely low momenta.
We counted the entries of the higher $dM$ peak as the $D^{*\pm}$
yield in the experiment,
and in the Monte-Carlo we corrected the peak entries
about -15\% considering the amount of the materials and their
solid angle from the interaction point. The error of this
correction was considered to be 5\% which was estimated from the
$e^+e^-$ conversion pair yield in the two photon sample.
Concerning the higher $dM$ fluctuation,
we have no explanations other than
statistical fluctuations.
The fitting function was the sum of a Gaussian and the following
 function:
 $$a(dM-M(\pi^\pm))^b(1-dM/c)^n,$$
 where a, b, and c are free parameters and n is an average multiplicity
 of the event sample.
The obtained $D^{*\pm}$ yield was $20.0\pm5.0$, where the $\chi^2$ of the
fit was 31 with DOF=35.
 The peak position and its width obtained by this fitting were
147.4 $\pm$ 0.2, and 0.8 $\pm$ 0.2 MeV, where the detector simulation
predicted 145.4 and 1.6 MeV, respectively. We considered that
the shift of the peak position was caused by the inhomogeneous
material distribution described above, whereas the energy
loss correction was carried out assuming uniform material.
The shift of the mass-difference peak quantitatively agreed with
the expectation. The resolution difference was due to the overestimation
of the material in front of the TPC in the simulation
in the most probable energy-loss case. 85\% of the soft-pions
were considered to pass away from the support structures of the TPC.
We tried $\chi^2$ and likelihood fitting and also tried polynomial
background functions. The differences in the total $D^{*\pm}$ yields
were within 10\%. Thus we concluded that the systematic error of the
fitting procedure was 10\%.
The yields for the four decay modes described before were
$9.5\pm3.3$, $4.8\pm2.3$, $4.8\pm2.2$, and $0.9\pm0.9$, respectively.

The background from $e^+e^-\rightarrow \gamma \rightarrow
q\bar{q}(\gamma)$ distributed smoothly in this $dM$ plot
(height was about 1
count a bin as shown in Fig.\ \ref{fig1}-(f)).
No peak structures were obtained by the Monte-Carlo
simulation. We also checked the event vertex distribution
in order to determine the contamination of the beam-gas background.
There was a clear peak at the interaction point with no tails.

\begin{figure}
\epsfysize7cm
\hskip1in\epsfbox{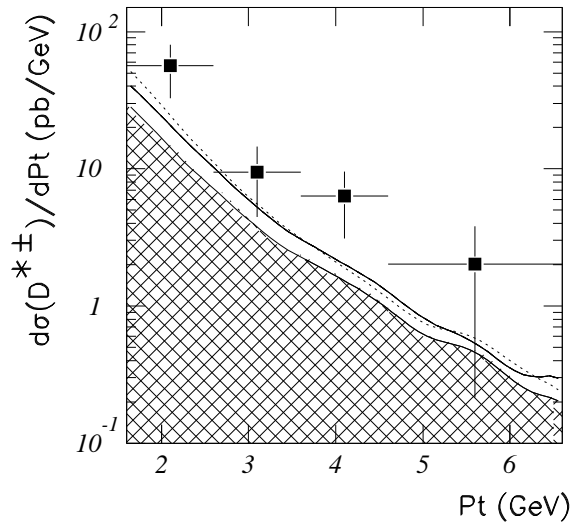}
\caption{
$d\sigma(e^+e^-\rightarrow e^+e^-
D^{*\pm}X)/dP_T$ versus $P_T$. The data
points with error bars are
the experimental data and the curves were obtained by Monte-Carlo
simulations while assuming various processes.
The hatched area is a prediction based on the direct process,
the solid curve
is based on a combination of direct and resolved photon
process by the GRV parametrization, the dotted one is based on
the LAC1 parameterization.
The vertical scales for the resolved photon process were
normalized by the relative acceptance to the direct process, because
of the acceptance difference described in the text.
}
\label{fig4}
\end{figure}

In order to compare the experimental data with the theoretical prediction,
we chose $d\sigma(D^{*\pm})/dP_T$,
instead of the total cross section, as had been
reported \cite{r7,r8,r9},
since the accepted $D^{*\pm}$ events were limited
to the high $P_T$ region due to the detector
acceptance.
We were sensitive to those $D^{*\pm}s$ with transverse momenta
$(P_T)$ between 1.6 and 6.6 GeV.
The lower limit was due to the detector acceptance and the higher
due to statistics.
An acceptance correction was
carried out by using a lowest-order (Born approximation) direct-process
Monte-Carlo simulation, in which
an equivalent photon approximation
was used.
The $P_T(D^{*\pm})$ smearing by the $P_T$ of the initial photons
was checked by a full calculation of the $e^+e^-\rightarrow
e^+e^-\gamma\gamma\rightarrow e^+e^-c\bar{c}$ process, and was
concluded to be small.
The hadronization was taken care of by
LUND6.3 Monte-Carlo \cite{r12,r22}.
We did not use a parton shower option in the LUND Monte-Carlo,
since the choice of $Q^2$ of the event was not well defined.
The fragmentation parameters in the calculation have
been tuned to reproduce the general properties of
single-photon exchange processes.
 The systematic ambiguity for $D^{*\pm}$ inclusive spectrum was
 considered to be at most 10 \%. Because the fragmentation of the $D^{*\pm}$
 was measured well at various $\sqrt{s}$ and the symmetric LUND
 fragmentation function reproduces it very well.
The systematic uncertainty of the cross section is mostly due to
the fitting procedure, fragmentation scheme (described so far),
the
branching ratios of $D^0$ \cite{r13} and $D^{*+}$ \cite{r14},
the ambiguities in the detector acceptance and the calibrations.
It was estimated to be 15\%.
The resulting differential cross sections are
$57\pm24$, $9.5\pm5.1$, $6.3\pm3.2$ and $2.0\pm1.8$ pb/GeV
(systematic errors included) for
$P_T$ regions 1.6-2.6 ,2.6-3.6, 3.6-4.6, and 4.6-6.6 GeV,
respectively,
and are plotted in
Fig.\ \ref{fig4}, together with the theoretical predictions.
The raw entries at each $P_T$ binnings were 7.2$\pm$2.9,
6.4$\pm$3.3, 4.0$\pm$2.0, and 2.4$\pm$2.1, where the errors
are statistical error only, respectively.

Theoretical predictions for both direct and resolved photon
processes
were compared with this measurement as follows.
In order to estimate $\sigma (D^{*\pm})$ from
the theoretical calculation of $\sigma _{c\bar{c}}$,
we assumed $P_{c\rightarrow D^{*\pm}}\times B(D^{*+}\rightarrow
\pi^+D^0)=0.185\pm0.024$ at $\sqrt{s}=90$ GeV in Ref.\ \cite{r15}.
This value is consistent with that obtained at $\sqrt{s}=10$ GeV
\cite{r25}.
The events were generated by the lowest-order calculations. The
relative acceptance for the process of the
resolved photon to that of the direct process
was obtained to be $0.76\pm0.25$ due to
the event selection, where the error was
especially due to the gluon-density function dependence.
Although, this ambiguity is small compared with the total contribution
including the direct process.
Since we did not measure the
energy flow in the low-angle region,
spectator jets were not detected.
This lower acceptance was especially due to the
lower visible energy cut of 4 GeV.
The vertical scale for the cross sections of the resolved
photon process shown in Fig.\ \ref{fig4} was normalized by this factor.

Corrections of order $\alpha_s$ were carried out
according to the procedure in Ref.\ \cite{r16}.
Since string fragmentation includes a parton-shower-like effect,
the next-to-leading-order correction in the $P_T$
spectrum is doubly counted. We used instead the $P_T$-independent
correction to the direct process.
The cross section of the direct process is
increased by a factor of 1.31 uniformly, and
that of the resolved photon process is corrected by a $P_T$
dependent function, due to the presence of
the process $\gamma q\rightarrow
c\bar{c}q$ (a part of this is absorbed in the gluon density
function in the resolved photon).
The $P_T$ dependent factors were obtained by the following way:
 At first we derived the $P_T$ dependent ratios between the higher
 and the lowest order calculations for the direct and resolved
 photon processes; we then calculated the factors between those of
 the direct and resolved photon process; and those factors were
 normalized to fit to the total cross section of the higher order
 calculations for the resolved photon process. The resulted $P_T$
 dependent correction were written as
  $$0.50P_T^c+0.54,$$
where $P_T^c$ is a $P_T$ of charm quark.
The charm quark mass ($m_c$) and the renormalization scale ($\mu$)
were assumed to be
$m_c$=1.6 GeV and $\mu=\sqrt{2} m_c$, respectively.
The curves in Fig.\ \ref{fig4}
represent these predictions. The parametrization
dependence for the resolved photon process appears in the lowest
$P_T$ binning.
Although our data favor GRV\cite{r5} or LAC1\cite{r4}, none
of them explain the high $P_T$ excess
(+1.5$\sigma$ at $P_T>$ 2.6 GeV).

There are ambiguities in the above-mentioned
theoretical predictions, which
depend on
$m_c$ and $\mu$. We changed $m_c$ from 1.3 to 1.8 GeV and $\mu$ from
$m_c/2$ to $2m_c$. The lower $m_c$ and $\mu$ give a higher cross section
which has a $P_T$ dependence. The lowest $P_T$ region may be
explained by such ambiguities as mentioned above, and the threshold
enhancement of $c\bar{c}$ production.
We thus concentrate on the higher $P_T$ regions,
i.e., $P_T>$2.6 GeV.
The case for $m_c=1.3$ GeV and $\mu=
m_c/2$ gives the highest one, i.e., +0.54pb (+6\%) higher than the nominal
value.
The experiment gives $\sigma(2.6<P_T<6.6~GeV)=
19.8\pm7.0pb$, where the GRV parametrization predicts $9.0\pm0.5$ pb.
Then the
difference is $10.8\pm7.5$pb.

In addition, if gluons inside the photon have
$P_Ts$ of 0.4 GeV in average, the $P_T$s of $c\bar{c}s$ are shifted.
This may increase the resolved
photon cross section by about a factor of two, because the spectrum
of this process is proportional to an exponential function. Therefore,
the $2\sim3$pb increase may be expected by this
(no increase in the direct process).
Then our results become consistent with the expectation within
1 $\sigma$ level.

Such event shapes as missing $P_T$ and thrust distributions were checked.
These shapes are consistent with the prediction of the lowest-order
direct and resolved photon processes, except for the overall
normalization factor.

In a part of the data set (integrated luminosity of 90 $pb^{-1}$),
there were forward calorimeters (FCL: made of BGO) which
covered the polar angle region between 3.2 and 13.6 degrees \cite{r21}.
We analyzed tagged events
by the FCL with the same analysis.
We observed the total $D^{*\pm}$ yield to be $3.5\pm3.6$,
where the lowest-order direct process predicts $1.0\pm0.4$ events.
We therefore need more statistics in order to observe tags from
electrons and positrons
as well as spectator jets.

In conclusion, we measured the inclusive $D^{*\pm}$ production
cross section in two-photon collision events
at the TRISTAN $e^+e^-$ collider.
The mean $\sqrt{s}$ was 57.16 GeV and the integrated
luminosity was 150 $pb^{-1}$.
We observed $20\pm5$ $D^{*\pm}$ in our data sample. The production
cross section in
the high $P_T$ region was
$\sigma_{D^{*\pm}}(2.6<P_T<6.6~GeV)=19.8\pm7.0pb$.
Comparisons with the theoretical prediction of
the direct process and the resolved photon
process were carried out.

We appreciate discussions with Prof. M. Kobayashi (KEK),
Dr. M. Drees (Univ. of Wisconsin), M. Kr\"amer (DESY), and
J. Zunft (DESY).
We thank the TRISTAN accelerator staff
for the successful operation of TRISTAN. We also thank
all engineers and technicians at KEK as well as
members of the collaborating
institutions: H. Inoue, N. Kimura,
K. Shiino, M. Tanaka, K. Tsukada, N. Ujiie,
and H. Yamaoka.

\end{document}